\documentclass[aps,prl,reprint,showpacs,superscriptaddress]{revtex4-1}
\usepackage{graphicx}
\usepackage{dcolumn}
\usepackage{bm}
\begin{document}


\title{Stabilizing transverse ablative Rayleigh Taylor like instability \\
       by using elliptically polarized laser pulses \\
       in the hole-boring radiation pressure acceleration regime}
\author{Dong Wu}
\email{wudongphysics@gmail.com}
\affiliation{Center for Applied Physics and Technology, Peking University, Beijing, 100871, China.}
\affiliation{Key Laboratory of High Energy Density Physics Simulation, Ministry of Education,
Peking University, Beijing, 100871, China. }
\author{C. Y. Zheng}
\email{zheng\_chunyang@iapcm.ac.cn}
\affiliation{Center for Applied Physics and Technology, Peking University, Beijing, 100871, China.}
\affiliation{Key Laboratory of High Energy Density Physics Simulation, Ministry of Education,
Peking University, Beijing, 100871, China. }
\affiliation{Institute of Applied Physics and Computational Mathematics, Beijing, 100088, China.}
\author{C. T. Zhou}
\affiliation{Center for Applied Physics and Technology, Peking University, Beijing, 100871, China.}
\affiliation{Key Laboratory of High Energy Density Physics Simulation, Ministry of Education,
Peking University, Beijing, 100871, China. }
\affiliation{Institute of Applied Physics and Computational Mathematics, Beijing, 100088, China.}
\author{X. Q. Yan}
\affiliation{Center for Applied Physics and Technology, Peking University, Beijing, 100871, China.}
\affiliation{Key Laboratory of High Energy Density Physics Simulation, Ministry of Education,
Peking University, Beijing, 100871, China. }
\author{M. Y. Yu}
\affiliation{Institute of Fusion Theory and Simulation, Zhejiang University, Hangzhou, 310027, China.}
\author{X. T. He}
\email{xthe@iapcm.ac.cn}
\affiliation{Center for Applied Physics and Technology, Peking University, Beijing, 100871, China.}
\affiliation{Key Laboratory of High Energy Density Physics Simulation, Ministry of Education,
Peking University, Beijing, 100871, China. }
\affiliation{Institute of Applied Physics and Computational Mathematics, Beijing, 100088, China.}
\date{\today}
\begin{abstract}
It is shown that the transverse Rayleigh Taylor like instability can be well stabilized 
by using elliptically polarized laser in the hole boring radiation pressure acceleration regime. 
The $\bm{J}\times\bm{B}$ effect of the laser will thermalize the local electrons and support a transverse diffusion mechanism of the ions, resulting in the stabilization of the short wavelength
perturbations, which is quite similar to the ablative Rayleigh Taylor instability in the initial confinement fusion research. 
The proper range of polarization ratio is obtained from a theoretical model for the given laser intensity and plasma density. The stabilization mechanism is well confirmed by two dimensional Particle-in-Cell simulations, and the ion beam driven by the elliptically polarized laser is more concentrated and intense compared with that of the circularly polarized laser.  
\end{abstract}
\maketitle


Recently, ion acceleration from the interaction of ultra-intense laser pulse with plasmas has attracted wide attention because of its broad applications, including producing high energy density matter, ion-fast ignition in laser fusion, tumor therapy and 
radiographing\cite{PhysRevLett.106.145002, PhysRevLett.100.225001, PhysRevLett.86.436, PhysRevLett.85.2945, PhysRevLett.89.175003, PhysRevLett.88.215006}. Almost all of these applications call for a high quality ion beam with large particle number, sharp energy spread and low divergence angle. Radiation pressure acceleration (RPA) is an potential scheme for generating high quality ion beams. 
According to the target thickness, usually there are two modes of RPA acceleration mechanism: light sail (LS) RPA for thin target\cite{PhysRevLett.103.024801, PhysRevLett.102.145002, PhysRevLett.105.155002, PhysRevLett.105.065002, PhysRevLett.105.065002,
PhysRevLett.100.135003, PhysRevLett.103.135001, PhyPla.16.044501, PhysRevLett.108.225002}
and hole boring (HB) RPA for thick target\cite{PhyPla.18.056701, PhysRevLett.102.025002, PhyPla.16.083130, PlaPhyConFus.51.024004, PlaPhyConFus.51.095006, PhyPla.14.073101, PhyPla.18.073101, PhyPla.16.033102, PhysRevLett.106.014801}. In particular, the HBRPA owns the intrinsic property for large particle number acceleration\cite{PhyPla.18.053108}. In HBRPA, the ponderomotive force drives the local electrons inward, resulting in a shock like double layer (DL) region with large electrostatic charge separation field. The latter could trap and reflect the ions initially located ahead of the DL, compressing and accelerating them like a piston. For a usually circularly polarized (CP) laser driven HBRPA, the DL oscillations would broad the energy spread of the accelerated ion beams\cite{PlaPhyConFus.51.024004}. Wu et al. proposed to use elliptically polarized (EP) laser to suppress the DL oscillations, generating high quality mono-energetic ion beams compared with that of CP laser\cite{ArXiv}. However the crucial issue, which is of fundamental influence to the RPA acceleration scheme, is the transverse Rayleigh Taylor like instability (RTI)\cite{PhysRevLett.105.065002,PhyPla.18.073106,PhysRevLett.108.225002}. The classical RTI can occur when a light fluid pushes or accelerates a heavy fluid, and this situation is quite similar to the RPA case, where the photons act as light fluid and plasmas as heavy fluid\cite{PhysRevLett.108.225002}. 
The RTI will break the target surface and terminate the acceleration process. However, unlike the 
classical RTI, the short wavelength ablative RTI in initial confinement fusion (ICF) research can be stabilized due to thermal smoothing of the perturbation or the transverse diffusion mechanism\cite{Phys.Fluids.28.3676,PhyPla.3.1402,PhyPla.113.690,PhyPla.3.2122,PhysRevLett.98.245001}. 

In this paper, we propose to use EP laser for stabilizing the transverse RTI in the HBRPA. The idea is quite similar to the stabilization mechanism for ablative RTI in the ICF research. Because of the $\bm{J}\times\bm{B}$ effect, the EP laser will thermalize those electrons
located within the DL region, and the high plasma temperature provides a fast transverse diffusion velocity 
($\sim\sqrt{T_eZ/m_i}$) of the ion particles, 
where $T_e$ is the plasma temperature, $Z$ is the ion charge number and $m_i$ is the ion mass. It is this transverse diffusion of
the ions that stabilize the transverse ablative RTI. During the characteristic time of the RTI, the diffusion range of the ions can overshoot the instability wavelength if a faster diffusion velocity is given, equally a higher plasma temperature or smaller polarization ratio $\alpha=a_z/a_y$ for EP laser. However, as expected, if $\alpha$ is too small ($\alpha=0$ for linearly polarized laser), the laser piston structure is totally destroyed. Thus, there should be a lower limit for the polarization ratio $\alpha$ to sustain the HBRPA process. Based on these ideas, the proper range of polarization ratio is obtained through a theoretical model for given the laser intensity and plasma density. This scheme is well confirmed by two dimensional (2D) Particle-in-Cell (PIC) simulations. 

In the HBRPA regime as shown in Fig.\ 1 (a)\cite{PhysRevLett.102.025002, PhyPla.16.083130}, the ponderomotive force drives the local electrons inward, resulting in a shock like DL region with large electrostatic charge separation field. The latter could trap and reflect the ions, compressing and accelerating them like a piston.
Because of the smaller mass of electrons, the accelerated ion beam is accompanied with an electron beam, keeping almost quasi-neutral. If we think of the ion and electron motions as a whole,
it is reasonable to neglect the electron inertia, and assume that the ponderomotive force acts on the ions directly. As the acceleration
process is limited within the DL region, which is about tenth of laser wavelength, this situation can be greatly simplified\cite{PhysRevLett.29.1429,PhysRevLett.99.065002,PhyPla.19.043104}.

\begin{figure}\label{1}
\includegraphics[width=8.50cm]{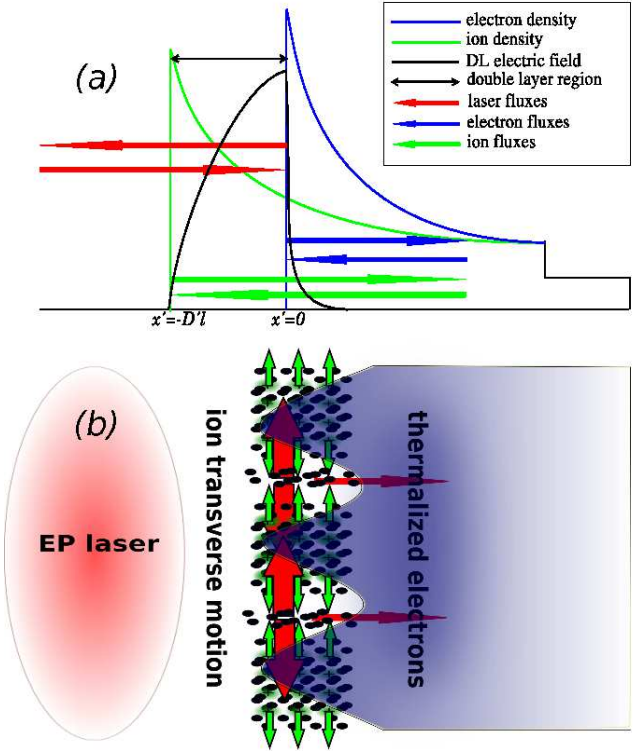}
\caption{\label{f1} (color online) (a) Schematic structure of the piston maintained by the radiation pressure in the piston-rest frame. (b) The stabilization mechanism of the EP laser driven ablative RTI. The EP laser thermalizes those electrons, 
providing a fast transverse diffusion velocity of the ion particles, which stabilizes the transverse ablative RTI.}
\end{figure}

Let us consider in the piston-rest frame, as shown in Fig.\ 1. The mass density of the thin layer is 
\begin{equation}\label{1}
\sigma_m =\int\limits_{-D^{\prime}_l}^{0}m_i n^{\prime}_idx^{\prime} \approx m_i n_i c/\omega_{pi},
\end{equation}
where $D^{\prime}_l$ is the DL width, $n_i^{\prime}$ is the ion density distribution within the DL region and $\omega_{pi}=\sqrt{4\pi Ze^{2}n_e/m_i}$ is the characteristic frequency of the ions.
The radiation pressure of the EP laser is 
\begin{equation}\label{2}
p_{rad}=2I_0(1-\beta_f)/(1+\beta_f)/c,
\end{equation}
where $I_0=(a_y^2+a_z^2)n_cm_ec^3/2$, $a_y$ and $a_z$ are the normalized laser amplitudes $a_y=eE_{y}/m_e\omega_{0}c$, $a_z=eE_{z}/m_e\omega_{0}c$, $\beta_f$
is the propagation velocity of the laser piston $\beta_f=\sqrt{I_{0}/n_i(m_i+Zm_e)c^{3}}/$ $[1+\sqrt{I_{0}/n_i(m_i+Zm_e)c^{3}}]$,
$m_e$ is the electron mass, $m_i$ is the ion mass, $Z$ is the ion charge number 
and the term $(1-\beta_f)/(1+\beta_f)$ is the modification due to the Doppler effect in the piston-rest frame. 
The acceleration $g$ of the inertial force acted on the thin layer of ions can be expressed as 
\begin{equation}\label{3}
g=p_{rad}/\sigma_m.
\end{equation}

\begin{figure}\label{2}
\includegraphics[width=8.50cm]{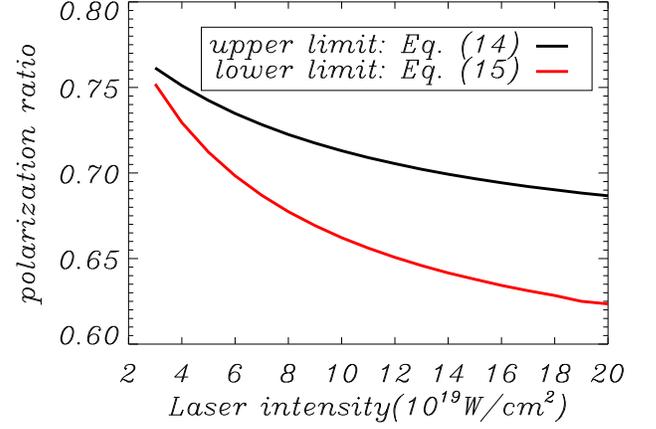}
\caption{\label{f2} (color online) The proper range of polarization ratio vs. laser intensity. Here, the laser wavelength is $10.0$ $\mu$m and the hydrogen plasma density is $20n_c$ which is $2.2\times10^{19}$ /cm$^{3}$. The black line represents the upper limit derived from Eq.\ (14) and the red line is for the lower limit from Eq.\ (15).}
\end{figure}

We consider two points $(x_0,y_0)$ and $(x_0,y_0+\delta y_0)$ on
the thin ion layer. These two points will evolve at some time to the points $(x,y)$ 
and $(x+\delta y_0\partial{x}/\partial{y_0},y+\delta y_0\partial{y}/\partial{y_0})$. The $x$ and $y$ components of the force equation of the element on the thin layer can be written as
\begin{eqnarray}\label{4,5}
&&\partial{p_x}/\partial{t}=-g\sigma_m d y_0+p_{rad}dy_0\partial{y}/\partial{y_0}, \\
&&\partial{p_y}/\partial{t}=-p_{rad}dy_0\partial{x}/\partial{y_0},
\end{eqnarray}
where $p_x=\gamma_f \sigma_m d y_0 dx/dt$ and $p_y=\gamma_f \sigma_m d y_0 dy/dt$. After some arrangement, 
Eq.\ (4) and (5) can be rewritten as
\begin{eqnarray}\label{6,7}
&&\partial^{2}{x}/\partial{t}^{2}=-g/\gamma_f+(\partial{y}/\partial{y_0})g/\gamma_f, \\
&&\partial^{2}{y}/\partial{t}^{2}=-(\partial{x}/\partial{y_0})g/\gamma_f.
\end{eqnarray}
The solution of Eq.\ (6) and (7) turns out to be\cite{PhysRevLett.29.1429}
\begin{eqnarray}\label{8,9}
&&x=\delta_0 \exp[t(kg/\gamma_f)^{1/2}]\cos(ky_0), \\
&&y=y_0-\delta_0 \exp[t(kg/\gamma_f)^{1/2}]\sin(ky_0),
\end{eqnarray}
where $\delta_0$ is the initial disturbed length, $k=2\pi/\lambda_{rt}$ and $\lambda_{rt}$ is the instability wavelength.
If the disturbed length in the $x$ direction is $1/k$ at sometime $\tau^{'}$, it means that the adjacent sections of the thin layer begin to collide with each other\cite{PhysRevLett.29.1429}. Here the transverse RTI has already developed, and we define $\tau^{'}$ as the characteristic time of the RTI. From Eq.\ (8), we have
\begin{equation}\label{10}
\tau^{'}=(\gamma_f/kg)^{1/2}\log(1/\delta_0k).
\end{equation}
When transformed to the laboratory frame, the characteristic time of the RTI reads, $\tau=\gamma_f\tau^{'}$,  
\begin{eqnarray}\label{11}
\tau=\frac{\lambda^{1/2}\gamma^{3/2}_f (m/Z)^{3/4}n^{1/4}(1+\beta_f)^{1/2}}{2\pi(a_y^2+a_z^2)^{1/2}(1-\beta_f)^{1/2}}\log(1/\delta_0k),
\end{eqnarray}
where $m=m_i/m_e$, $n=n_e/n_c$, $n_c$ is the critical density, $\lambda=\lambda_{rt}/\lambda_0$ is the instability wavelength normalized to laser wavelength, and $\tau$ is normalized to laser period $T_0$.

The factor $\log(1/\delta_0k)$, to some degree, is constant, which can be determined from the PIC simulations.
It must be emphasized that $\log(1/\delta_0k)$ is not sensitive to $\delta_0$ and $k$. Once the value of $\log(1/\delta_0 k)$ is chosen, it will keep constant for all cases of different laser intensity and plasma density.  
According to our experience of a series of 2D PIC simulations, we have already defined the constant factor $\log(1/\delta_0k)=13.8$ compared with Eq.\ (11). This equation demonstrates that the shorter wavelength perturbations grow more faster, and that strong laser intensity and low plasma density correspond to short characteristic time of instability.

When an EP laser falls on the target, its oscillating ponderomotive force will thermalize the local electrons. Considering the 
electron gamma factor under the EP laser, 
\begin{equation}\label{12}
\gamma_e=\sqrt{1+[a_y\cos(\omega_0t)]^2+[a_z\sin(\omega_0t)]^2},
\end{equation} 
and assuming $1+a_y^2+a_z^2>>a_y^2-a_z^2$, Eq.\ (12) can be transfered to, using Taylor expanding formula,
\begin{equation}\label{13}
\gamma_e=\gamma_{ec}+\gamma_{eos}\cos(2\omega_0t),
\end{equation}  
where $\gamma_{ec}=\sqrt{1+a_y^2/2+a_z^2/2}$ is the constant gamma term and $\gamma_{eos}=(a_y^2-a_z^2)/4\gamma_{ec}$ is the oscillating gamma term.
The thermalization of the local electrons is dominated by the time oscillating term, in which the temperature of the local electrons can be approximated to $m_ec^2(\gamma_{eos}-1)$. Thus, the transverse diffusion velocity of ions, equally the sound speed, is $v_d=\sqrt{(\gamma_{eos}-1)Z/m}$, where $v_d$ is normalized to the light speed $c$.

As shown in Fig.\ 1 (b), in order to stabilize the transverse ablative RTI, we suppose that during the characteristic time $\tau$, the diffusion range of the ions must overshoot the instability wavelength\cite{PhyPla.18.073106}. To smooth the short wavelength perturbations with wavelength  $\lambda<1$, we have $\tau v_d > 1$. After some arrangement, it reads
\begin{widetext}
\begin{eqnarray}\label{14}
(a_y^2-a_z^2)/4\sqrt{1+a_y^2/2+a_z^2/2}-1 > \eta(a_y^2+a_z^2)(1-\beta_f)/(m/Z)^{1/2}\gamma_f^3n^{1/2}(1+\beta_f),
\end{eqnarray}
\end{widetext}
where $\eta=[2\pi/\log{(1/\delta_0k)}]^{2}=0.21$.
From Eq.\ (14), the minimum temperature equally the upper limit of the polarization ratio $\alpha$ is determined for the defined laser intensity and plasma density.

There must be, as expected, a lower limit for the polarization ratio $\alpha$ to keep the piston structure intact. For an usually laser piston structure shown in Fig.\ 1 (a), the ponderomotive force drive the local electrons inward resulting in a shock like DL region with large electrostatic charge separation field, which could trap and reflect the ions. While the strong $\bm{J}\times\bm{B}$ effect of LP laser will drag the local electrons forward to vacuum breaking the laser piston structure.
The lower limit of the polarization ratio $\alpha$ is derived based on the assumption that these forward-going electrons must be stopped within the DL region. Balancing the (forward) kinetic and the electrostatic and ponderomotive potential energies of the electrons\cite{ArXiv}, we can obtain the lower limit of the polarization ratio $\alpha$, 
\begin{widetext}
\begin{eqnarray}\label{15}
m(\gamma_f-1)/Z+\sqrt{1+(a^2_y-a^2_z)^2/16}-1 < \sqrt{m/Z}(a^2_y+a^2_z)\beta_f(1-\beta_f)/3(1+\beta_f)\gamma_fn.
\end{eqnarray}
\end{widetext}
For the given laser intensity $I_0=(a_y^2+a_z^2)n_cm_ec^3/2$, plasma density $n_e$, ion mass $m$ and ion charge number $Z$, we can solve out two polarization ratios $\alpha=a_z/a_y$ from Eq.\ (14) and (15). These two values are the upper and lower limit of suitable polarization ratios.  
For the hydrogen plasma with density $n_e=20n_c$, 
the proper range of polarization ratio $\alpha=a_z/a_y$ can be obtained according to 
Eq.\ (14) and (15), which is shown in Fig.\ 2.

\begin{figure*}\label{3}
\includegraphics[width=16.0cm]{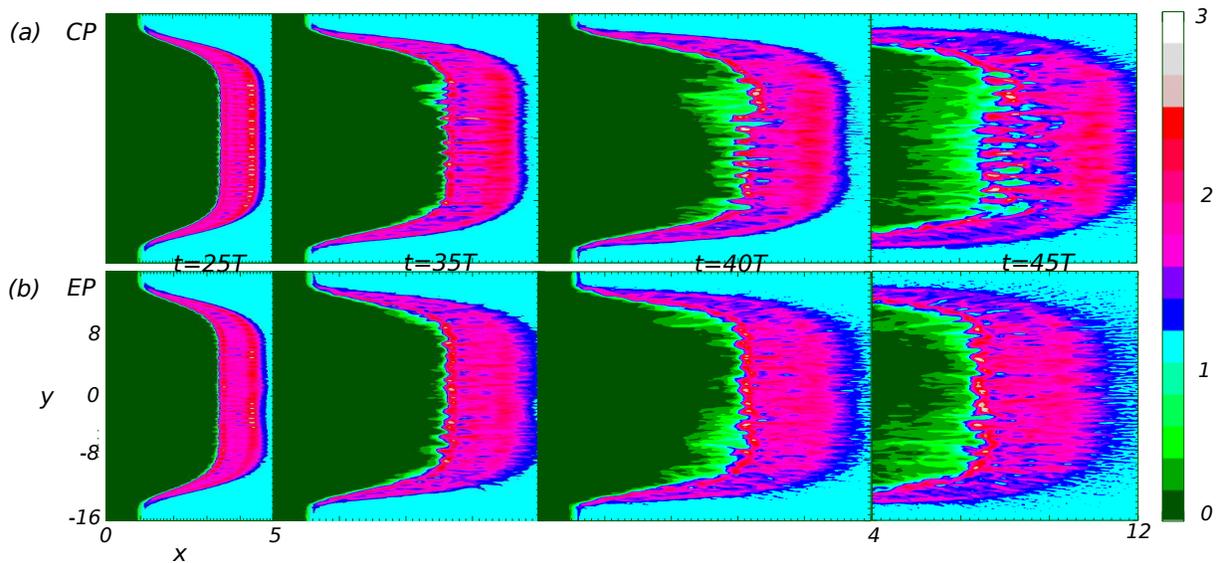}
\caption{\label{f3} (color online) (a) and (b) The distribution of proton density driven by CP and EP laser pulse at $t=25T_0$, $t=35T_0$, $t=40T_0$ and $t=45T_0$ respectively. Here, the laser wavelength is $10.0$ $\mu$m, the hydrogen plasma density is $20n_c$ which is $2.2\times10^{19}$ /cm$^{3}$, the laser intensity is $6.85\times10^{19}$ W$/$cm${^2}$ and the polarization ratio $\alpha=0.70$ for the EP laser pulse.}
\end{figure*}

2D PIC (\texttt{KLAP} code\cite{Eur.Phys.Lett.95.55005,PhysRevLett.107.265002}) simulations are run to further confirm this scheme. The size of the simulation box is $L_z\times L_y=20\lambda_0(z)\times40\lambda_0(y)$ with $\lambda_0$ representing the laser wavelength. The simulation box is divided into uniform grid of $2000(z)\times4000(y)$. The CP laser pulse enter into the simulation box from the left boundary. The bulk target
consists of two species: electrons and photons, which are initially located in the region $1.0\lambda_0<z<20.0\lambda_0$ and
$-20.0\lambda_0<y<20.0\lambda_0$ with density $n_e=20n_c$, where $n_c=\omega_0^2e^2m_e/4\pi=1.1\times10^{19}$ /cm$^3$ is the critical density for $10.0$ $\mu$m laser pulse \cite{PhysRevLett.106.014801,Nat.Phys.8.95}. We use $160$ particles per cell to run the simulations. The initial plasma temperature is set to be $10.0$ KeV. To exclude the effect like electron rebound on the target boundary, we applied particle absorbing boundary conditions, where the absorbed particles are compensated by the incoming particles with the background temperature from the boundary to ensure charge neutrality. The normalized amplitude of the CP laser electric field is $a_y=50.00$ and $a_z=50.00$, corresponding to the laser intensity $6.85\times10^{19}$ W$/$cm${^2}$. To exclude the effect of target surface curvature caused by the shape of the laser pulse, we applied transverse super-Gaussian laser pulse. The laser pulse has a temporally Gaussian  rising profile $\exp[-t/25)^2]$ followed by a constant intensity, where $t$ is normalized to laser period $T_0$. In contrast, the EP laser pulse with the same space and temporal profile is also run. According to Fig.\ 2, for laser intensity at $6.85\times10^{19}$ W$/$cm${^2}$, we choose the polarization ratio $\alpha=0.7$, which equally corresponding to $a_y=57.93$ and $a_z=40.55$.     

\begin{figure}\label{4}
\includegraphics[width=8.5cm]{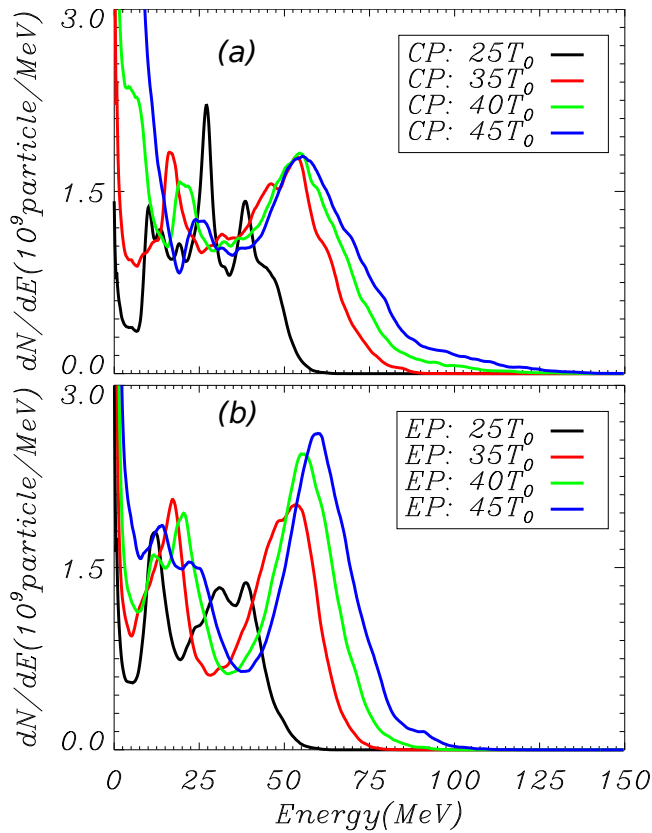}
\caption{\label{f4} (color online) (a) and (b) The energy spectrum of the proton beams driven by CP and EP lasers
at $t=25T_0$, $t=35T_0$, $t=40T_0$ and $t=45T_0$ respectively. Here, the laser wavelength is $10.0$ $\mu$m, the hydrogen plasma density is $20n_c$ which is $2.2\times10^{19}$ /cm$^{3}$, the laser intensity is $6.85\times10^{19}$ W$/$cm${^2}$ and the polarization ratio $\alpha=0.70$ for the EP laser pulse.}
\end{figure}

Our main results are clearly shown in Fig.\ 3. When driven by CP laser, the ponderomotive force has no second-harmonic oscillating component, and there is no thermalization effect at all. As shown in Fig.\ 3 (a) the transverse RTI has rapidly developed at $t=40T_0$, and the perturbations turn out to be even more fierce at $t=45T_0$, with the perturbation wavelength about $1\sim2\lambda_0$. These short wavelength perturbations can severely disturb the hole boring process, terminating the HBRPA acceleration at early time.  
In contrast, when driven by EP laser, because of the $\bm{J}\times\bm{B}$ effect, the EP laser will thermalize those electrons
located within the DL region, 
and the high plasma temperature provides a fast transverse diffusion velocity ($\sim\sqrt{T_eZ/m_i}$) of the ion particles. 
In this case, the temperature of the local thermalized electrons can be as high as $3$ MeV, thus the diffusion velocity can reach as fast as $v_d=0.05$. It is this transverse diffusion of the ions that stabilize the transverse ablative RTI, which is clearly shown in Fig.\ 3 (b). 
During the characteristic time of the RTI, $\tau\sim40T_0$, the diffusion range of the ions can be as far as $2.0\lambda_0$. 
The range of the transverse diffusion ions can readily overshoot the perturbations of short wavelength about $1\sim2\lambda_0$. In addition, as expected, the laser piston structure is kept intact. The HBRPA process sustains for a rather long time compared with that of CP laser. 

\begin{figure}\label{5}
\includegraphics[width=8.5cm]{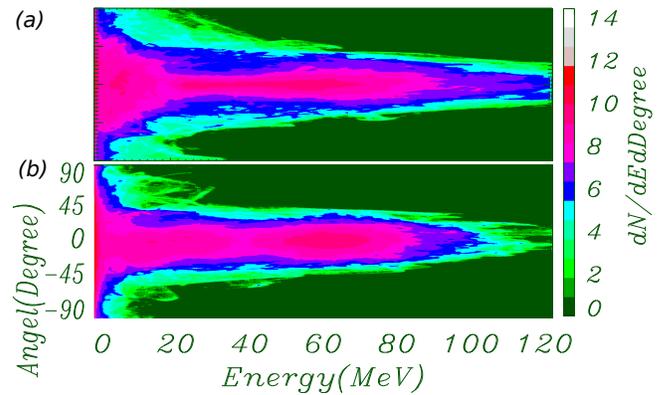}
\caption{\label{f5} (color online) (a) and (b) The angular distributions of the proton beams driven by CP and EP lasers respectively at $t=45T_0$. }
\end{figure}

The energy spectrum of the proton beam driven by the EP laser is much better than that of the CP laser, because of the relatively smooth laser piston surface or shock like electrostatic charge separation field, which is clearly shown in Fig.\ 4.
For the EP laser, because of the stabilization of the transverse RTI, the number of the accelerated proton particles is always accumulating resulting from the stable HBRPA process. While for the CP laser, the accumulating process is terminated at $t=40T_0$, 
and the chaotic heating is performed instead. 

Fig.\ 5 shows the angular distribution of the accelerated proton beams by CP and EP lasers respectively: (a) is for CP laser and (b) for EP. Compared with (a) and (b), we conclude that the transverse diffusion mechanism driven by EP laser does not have a obvious affection on the angular distribution of the proton beam. The transverse diffusion velocity of protons in this situation is around $v_d=0.05$, while the longitudinal velocity is double the piston forward velocity, which can be as high as $v_l=0.4$. It is reasonable to say that the small divergence angel can still be held under the mechanism we proposed. 
It demonstrates that the divergence angels of proton beams driven by both CP and EP lasers can be maintained within $10$ degree. However, the proton beam driven by EP is more intense and concentrated than that of CP. 

It should be emphasized that different polarization ratios with $0.65<\alpha<0.75$ are also run. The stabilization mechanism is not quite sensitive to the polarization ratio. The scheme is confirmed to be rather robust. 

In summary, we propose to use EP laser to stabilize the transverse RTI in the 
hole boring radiation pressure acceleration regime. The $\bm{J}\times\bm{B}$ effect of the laser will thermalize the local electrons and support a transverse diffusion mechanism of the ions, resulting in the stabilization of the short wavelength
perturbations, which is quite similar to the ablative Rayleigh Taylor instability in the initial confinement fusion research. 
The proper range of polarization ratio is obtained from a theoretical model for the given laser intensity and plasma density. The stabilization mechanism is well confirmed by two dimensional PIC simulations, the ion beam driven by elliptically polarized laser is more concentrated and intense compared with that of circularly polarized laser.  
\begin{acknowledgments}
One of the authors Dong Wu thanks B. Qiao for fruitful discussions.
This work was supported by the National Natural Science Foundation of
China (Grant Nos. 11075025, 10835003, 10905004, 11025523, and 10935002) and the Ministry of Science and Technology of China (Grant No. 2011GB105000).
\end{acknowledgments}

{}

\end{document}